\documentclass{vgtc}                           
\newcommand*{\doi}[1]{\href{http://dx.doi.org/#1}{doi: #1}}

\ifpdf%
\pdfoutput=1\relax              
\usepackage{graphicx}           
\DeclareGraphicsExtensions{.pdf,.png,.jpg,.jpeg} 
\else%
\usepackage{graphicx}                
\DeclareGraphicsExtensions{.eps}     
\fi%

\graphicspath{{figures/}{pictures/}{images/}{./}} 

\usepackage{arydshln}

\usepackage{microtype}                
\PassOptionsToPackage{warn}{textcomp}  
\usepackage{textcomp}                  
\usepackage{mathptmx}                  
\usepackage{times}                     
         
\usepackage{cite}                      
\usepackage{tabu}                      
\usepackage{booktabs}    
\usepackage{array}
\usepackage{pifont}
\usepackage{tikz}

\usepackage{color}
\usepackage{graphicx}
\usepackage{capt-of}

\usepackage{wasysym}

\usepackage{amssymb}

\usepackage{xspace}

\usepackage{enumitem}

\usepackage{hyperref}
\hypersetup{
    unicode=false,          
    pdftoolbar=true,        
    pdfmenubar=true,        
    pdffitwindow=false,     
    pdfstartview={FitH},    
    pdftitle={My title},    
    pdfauthor={Author},     
    pdfsubject={Subject},   
    pdfcreator={Creator},   
    pdfproducer={Producer}, 
    pdfkeywords={keyword1, key2, key3}, 
    pdfnewwindow=true,      
    colorlinks=true,       
    linkcolor=black,          
    citecolor=black,        
    filecolor=magenta,      
    urlcolor=black           
}

\usepackage{tabu}

\usepackage{picinpar}

\usepackage[utf8]{inputenc}

\usepackage{microtype}
\usepackage{wrapfig}

\usepackage{subcaption}

\usepackage{hyphenat}

\usepackage[english]{babel}
\addto\extrasenglish{%
}

\providecommand*\comment[1]{}
\let\commentstart=\iffalse

\providecommand*\thisisacomment[1]{}
\let\thisisacommentstart=\iffalse

\newcommand{\subhead}[1]{\vspace{2pt} \noindent \textbf{#1}}

\newcommand*\annotatedFigureText[4]{\node[draw=none, anchor=south west, text=#2, inner sep=0, text width=#3\linewidth,font=\sffamily] at (#1){#4};}
\newenvironment {annotatedFigure}[1]{\centering\begin{tikzpicture}
\node[anchor=south west,inner sep=0] (image) at (0,0) { #1};\begin{scope}[x={(image.south east)},y={(image.north west)}]}{\end{scope}\end{tikzpicture}}

\onlineid{10}

\vgtccategory{Research}

\vgtcpapertype{Position Paper}

\usepackage{color}
\usepackage{todonotes}
\usepackage{menukeys}

\definecolor{data}{HTML}{6d2c20}
\definecolor{analytic}{HTML}{757f24}
\definecolor{domain}{HTML}{3a3f91}
\definecolor{visual}{HTML}{217754}

\newcommand\data[1]{\textcolor{data}{\textit{#1}}}
\newcommand\analytic[1]{\textcolor{analytic}{\textit{#1}}}
\newcommand\domain[1]{\textcolor{domain}{\textit{#1}}}
\newcommand\visual[1]{\textcolor{visual}{\textit{#1}}}

\newcommand\acc[1]{\hspace{0.02cm},\hspace{-0.1cm}#1}

\title{Framing Visual Musicology through Methodology Transfer}

\DeclareMathSymbol{\ast}{\mathbin}{symbols}{"03}

\author{
Matthias Miller\thanks{e-mail:lastname@dbvis.inf.uni-konstanz.de}, ~
Hanna Sch\"afer\acc{$^*$} ~
Matthias Kraus\acc{$^*$} ~
Marc Leman\thanks{e-mail:marc.leman@ugent.be},  ~
Daniel Keim\acc{$^*$} ~
Mennatallah El-Assady$^*$
\\\scriptsize {$^*$}University of Konstanz \hspace{4cm} {$^\dag$}Ghent University 
}

\shortauthortitle{? \MakeLowercase{\textit{et al.}}: Framing Visual Musicology through Methodology Transfer}

\abstract{
In this position paper, we frame the field of \textit{Visual Musicology} by providing an overview of well-established musicological sub-domains and their corresponding analytic and visualization tasks. 
To foster collaborative, interdisciplinary research, we discuss relevant data and domain characteristics. We give a description of the \textit{problem space}, as well as the \textit{design space} of musicology and discuss how existing problem-design mappings or solutions from other fields can be transferred to musicology. 
We argue that, through \textit{methodology transfer}, established methods can be exploited to solve current musicological problems and show exemplary mappings from analytics fields related to text, geospatial, time-series, and other high-dimensional data to musicology.
Finally, we point out open challenges, discuss research gaps, and highlight future research opportunities.

} 

\keywords{Visual musicology, methodology transfer, visualization, music analysis, design space, research opportunities}

\teaser{
	\centering
	\vspace{-5pt}
    \begin{annotatedFigure}
        {\includegraphics[width=0.9\linewidth]{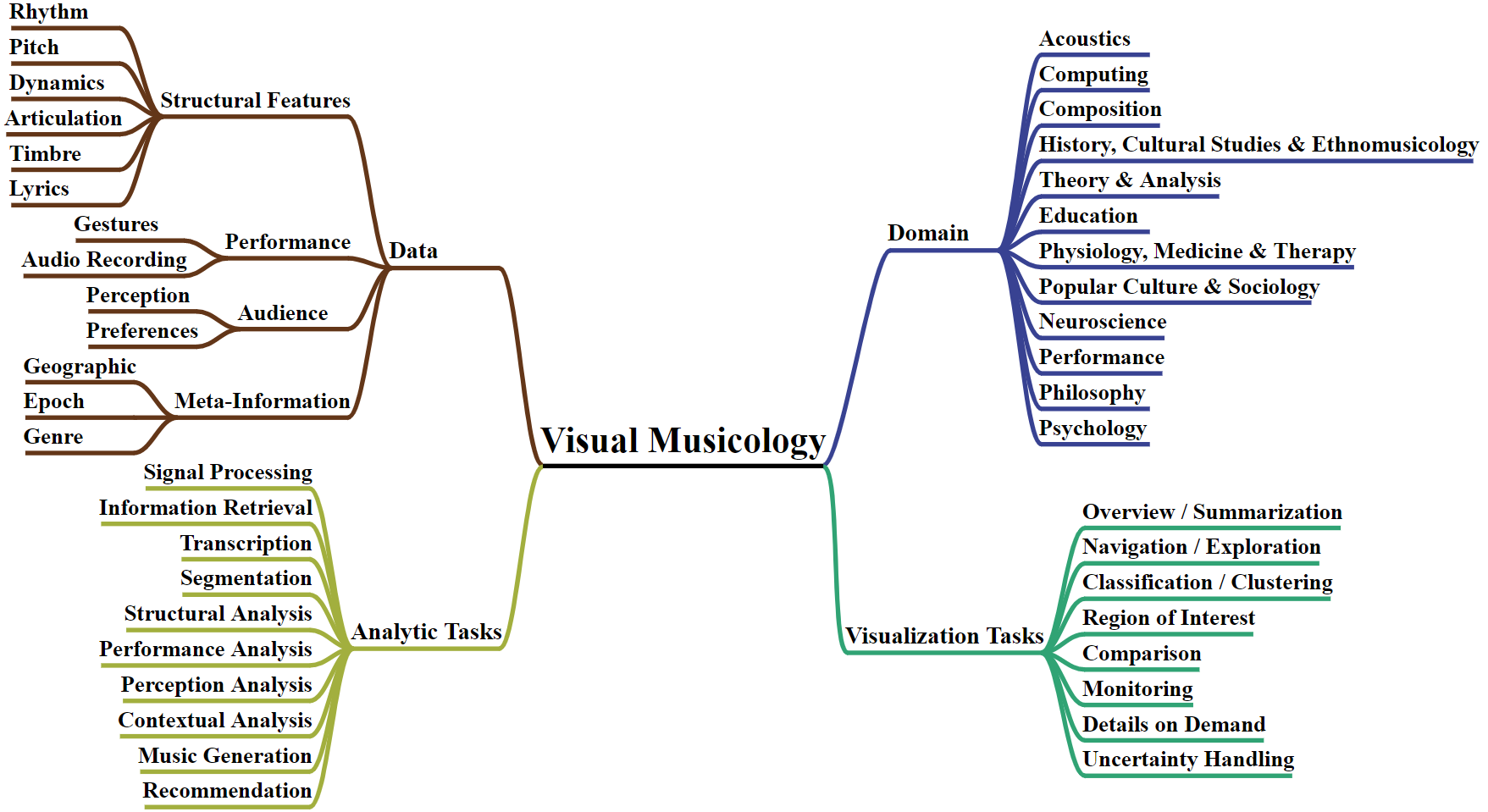}}
	\annotatedFigureText{0.2,0.95}{black}{0.51}{
	\begin{center}	    \small \hspace*{-1.1cm}\texttt{\href{https://visual-musicology.com/graph/}{https://visual-musicology.com/graph/}}	
	\end{center}}
    \end{annotatedFigure}
\caption{ 
Visual Musicology 
encompasses different \textit{data} types and various \textit{domains} covering psychological, physical, physiological, and societal aspects. 
	\textit{Visualization tasks} can support musicologists in answering domain problems, which can be described by  \textit{computational tasks}.  
	Strengthening cross-disciplinary and domain-driven research has the potential for original and synergistic research, paving the way for novel research collaborations between the musicology and information visualization domains.
	}
	\label{fig:teaser}
}

\usepackage{tikz} 
\usetikzlibrary{calc}
\usepackage{wasysym}
\begin{document}

\firstsection{Introduction}
    \maketitle
With its long history, music is an omnipresent element of cultures and a primary aspect of societal identity, deeply rooted in psychology, art, and entertainment~\cite{herndon1981music,whiteley_music_2005}. The current technological progress has a strong influence on music research, such as the application of new methods for optical music recognition~\cite{DBLP:journals/ijmir/RebeloFPMGC12}, music recommendation systems (e.g. Spotify~\cite{DBLP:conf/gvd/PichlZS14}), and music information retrieval~\cite{DBLP:journals/csr/KaminskasR12}.

As a subfield of the humanities, \textit{musicology} comprises any research-based analysis or study which is related to music~\cite{beard_musicology2005}.
In musicology, not only basic musical elements (e.g., rhythm or harmony) are of interest, but also \textit{ethnomusicological} considerations including gestures and movement~\cite{beard_musicology2005}. Music  itself is only one aspect within musicology. Musicologists study a variety of topics such as genres~\cite{aucouturier_representing_2003}, epochs~\cite{cook_musical_2013}, composers~\cite{mcandrew_music_2015}, and music theory~\cite{beard_musicology2005}, but also psychological effects and emotions~\cite{chanda_neurochemistry_2013}. 
Often, vast amounts of data and complex relationships within them require computational solutions to enable domain experts to extract interesting patterns. 

To achieve this, Visual Analytics (VA) help in automatic extraction and processing of valuable information while considering the needs of analysts~\cite{DBLP:series/lncs/KeimAFGKM08}. 
Incorporating the computational power enables us to deal with the complexity of large musicological data sources~\cite{watson_musescore_2018}. E.g., using large data sources increases the quality of music transcription~\cite{DBLP:journals/jiis/BenetosDGKK13} or improves music retrieval and recommendation~\cite{DBLP:journals/csr/KaminskasR12}.
%
In the information visualization community, many published approaches support domain-specific challenges, especially for analyzing text~\cite{DBLP:conf/apvis/KucherK15}, time~\cite{DBLP:series/hci/AignerMST11}, geography~\cite{kraak_cartography_2013}, and high-dimensional data~\cite{liu2016visualizing}. McNabb~\cite{DBLP:journals/cgf/McNabbL17} provides an overview of all surveys conducted by the visualization community until 2017. A search by the keywords \textit{music}, \textit{sound}, \textit{audio}, or \textit{harmony} using the SoS Literature Browser yields no results~\cite{sos_literature_browser_mcnabb_2017}. Still, a few individual applications exist, such as the web service proposed by Goto et al.~\cite{DBLP:conf/ismir/GotoYFMN11} for automatic chord extraction to support active music listeners and musicians by utilizing volunteered information to improve the automatic results. 

Thus, we assume that, compared to other domains, music visualization and visual analytics is an emerging and under-researched field with many open research opportunities. 
In this paper, we propose the term \textit{Visual Musicology} to describe the interdisciplinary research field at the intersection of musicology and visual analytics.
We argue that visual musicology will strongly benefit from transferring existing visualization techniques. Visualization researchers can contribute to unsolved challenges of the music domain, and the uniqueness of musicology offers them the opportunity to develop new visualization techniques and applications~\cite{wrisley_visualization_2018}. 


This position paper explains how to identify direct or indirect similarities between the music domain and other established visualization research domains, including text, geo, time, and high-dimensional data. Based on this knowledge, we introduce how to apply \textit{methodology transfer} to map existing visual solutions to benefit from previous research work. Moreover, we recommend developing new music visualization techniques to cover current research gaps. We propose ``\textit{Visual Musicology}'' to encapsulate information visualization research that addresses any aspect of music or musicology. It encompasses all design models and techniques in the visualization domain that provide a solution for problems related to musicology. 

For our contributions, we first characterize the \textit{problem} and the \textit{design space} of \textit{visual musicology} similar to TextVis~\cite{DBLP:journals/computers/AlharbiL19}. Second, we describe parallels between the musicology problem space comprising data characteristics, users, domains, and goals and problem spaces of other application domains. Third, we divide the \textit{visual musicology design space} into existing solutions and solutions that can be transferred or adapted from other domains, thus highlighting research gaps. Fourth, we conduct the \textit{methodology transfer} process by presenting three exemplary use cases to emphasize the potential of transferring existing visual solutions.

We postulate that collaboration between musicologists and visualization researchers may lead to mutually beneficial synergies: Visualization researchers can design and evaluate new methods by solving complex and diverse tasks of musicologists. Musicologists can address previously unsolved questions in a new way. Through this position paper, we want to motivate and initiate this kind of collaboration.

\thisisacomment{
\todo{Check for missing bullet points}    
\begin{itemize}
\item Music is omnipresent, long history, represents culture and identity, large profitable market, deeply rooted in psychology, art, entertainment, etc. 
\item As a musician or musicologist, I want to study music about different questions. Analyze styles, epochs, genres, composers, psychological effects, mathematical relations in music, a theory of music/composition. 
\item VA offers an opportunity to extract the information needed for such analyses which would not be possible with the current manual methods. At the same time, automatic extraction of these patterns and correlations is not possible yet due to the complexity of the interdependencies
\item VA knowledge generation pipeline - Human+Machine FTW 
\item Many existing visualization techniques/methods explored and established in other domains (Text, Time, Bio, Geo, HD)
\item The domain of music is not explored and investigated as other fields as text and biological data~\. Until 2017, McNabb presents an overview of all existing surveys in the information visualization domain~\cite{DBLP:journals/cgf/McNabbL17}. A search using the keywords ``music", ``sound," ``audio," or ``harmony" in the SoS Literature Collection browser yields no results. Consequently, this research area holds potential for much of research work.
\item Music is yet another domain that can benefit from available techniques
\item Unique selling point of Music Data for visualization experts - Variety of data types (sequence, hierarchy, high dimensionality, text) included into different types of application (Mathematical vs. Creative)
\item large open datasets available (e.g., MuseScore (NNN sheets) youtube (millions of sound files e.g., Songle\cite{DBLP:conf/ismir/GotoYFMN11}), musicxml
\item Some existing methods can be directly transferred from other domains to music 
\item Some methods from other domains could be adapted to also fit to music data
\item Some unique features and challenges of music data offer the chance to develop new methods for visualization
\item This paper uses a methodology transfer to identify direct parallels, similarities, and transfer gaps between music visualization and already established visualization domains (geo, text, time, HD)
\item The contributions of this paper are: First, we span the problem and design space of musicology, similar to (cite textvis, cite timevis). Second, we describe parallels between the problem space (data, domain, user, goals) in musicology and other application domains. Third, we classify the visual musicology design space into existing or fitting solutions, adaptable solutions from other domains, and research gaps. Fourth, we describe the methodology transfer process through three use cases from different domains, thus showing the potential of existing solutions. 
\end{itemize}
}
\section{Background}
\label{background}

\subhead{Musicology} comprises any research effort covering aspects of music~\cite{kerman2009contemplating}. 
Thus, not only analysis on performance data such as audio, sheet music, or digital symbolic data as MusicXML~\cite{good_musicxml_2001} and MIDI are of interest. 
Moreover, psychological aspects such as emotions of music audiences as well as gestures and body movement analysis of performing musicians are relevant research areas within the realm of musicology. Consequently, musicology covers a diverse range of application domains that we state in \autoref{fig:teaser}. We emphasize that the ``Visual Musicology'' graph in \autoref{fig:teaser} provides a diverse list of domains that we extracted from the Journal of Interdisciplinary Music Studies~\cite{jims2019}. The domain list is subject to extension since there are many specialized subdomains which we cannot consider within the scope of this work. For instance, besides music data, meta-information such as the places where composers lived and worked as well as how the style of different composers influenced each other are typical tasks in the sub-domain of \textit{history} or \textit{cultural studies}.
Nevertheless, the diversity of application domains leads to many complex challenges that require tailored solutions to cope with them. Not every sub-domain of musicology requires analytical techniques or is concerned about music experience~\cite{agawu_analyzing_1997}. While visualization is not a solution for every problem in the music domain, we argue that there is still a vast amount of open research gaps that wait to be addressed by interdisciplinary researchers.

\subhead{Music Visualization} often is only concerned with analyzing structural features~\cite{weiss2014quantifying} in audio or performance data~\cite{widmer_computational_2004} or on the notation level~\cite{miller_analyzing_2018}. Using information visualization methods and models for music data is not a new endeavor. More than two decades ago, Smith and Williams proposed a different way for music notation to visualize music~\cite{DBLP:conf/visualization/SmithW97}.
Many alternative music notations have been proposed to deal with the legibility of the common music notation which primarily consists of the features \textit{rhythm}, \textit{harmony}, \textit{dynamics}, \textit{timbre}, and \textit{articulation}~\cite{miller_analyzing_2018}. Miller et al. analyzed multiple mappings in music notation from visual variables to musical features~\cite{miller_analyzing_2018}. Besides alternative visual mappings of musical features, harmony~\cite{sapp2001harmonic} and structure analysis~\cite{wattenberg2002arc} are often the subject of visualization models~\cite{tymoczko_geometry_2006,lerdahl_overview_1983}. While the more obvious visualization models in the music domain are mainly rooted in direct features of music data, there are many more music-related topics encompassed by \textit{musicology}. Additionally, models were often blindly applied without clear musicological motivation or goals during the early stages of computational musicology~\cite{Tzanetakis2007}. At the Dagstuhl seminar ``Multimodal music processing'' (11041) many topics regarding music processing of sheet and audio data were discussed by leading musicologists~\cite{muller_multimodal_2012}. One of the results was a recommendation for interdisciplinary research efforts to improve the understanding of complex relationships between music and its effect on the human mind or body.  Therefore, bringing music domain experts and information visualization experts into a conversation may yield further fruitful collaborations~\cite{DBLP:conf/vissym/SimonMKS15}, resulting in useful solutions for both sides. 
\section{Methodology Transfer}
\label{sec:methodology}
\begin{figure}
    \centering
    \includegraphics[width=\linewidth]{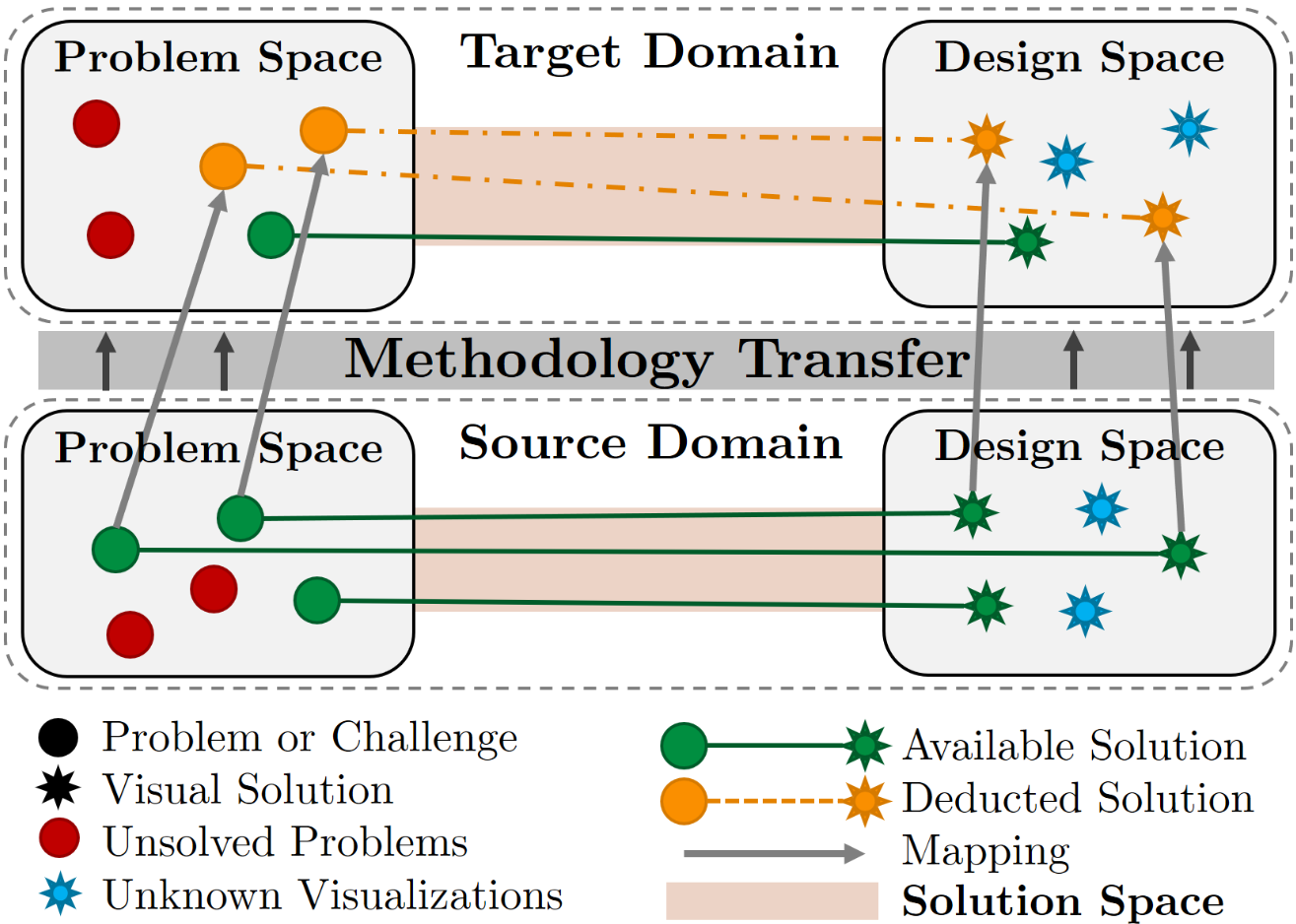}
    \caption{The Methodology Transfer Model (MTM) maps problems and designs between two different domains to deduce new solutions.  
    }
    \label{fig:method_transfer}
    \vspace{-5pt}
\end{figure} 
\textit{Methodology Transfer} refers to the action of utilizing available models that provide solutions to existing and unsolved problems~\cite{DBLP:conf/iv/Burkhard04}. We propose a \textit{Methodology Transfer Model (MTM)} which states that every domain contains a \textit{problem space} and a \textit{design space}, as depicted in \autoref{fig:method_transfer}.  Simon et al.~\cite{DBLP:conf/vissym/SimonMKS15} explain that problem spaces are defined by the \textit{data} and \textit{task} characteristics as well as \textit{domain-specific user needs}. The design space comprises all visual techniques that serve as a solution for issues from the problem space. We consider the mapping between these two spaces as the \textit{solution space}. Visual models from the design space may be applied to different problems. At the same time, a single visualization may cover various tasks from the problem space. To cope with new challenges, we need to find still unknown problem-design mappings which are located in the \textit{solution space}. Then, they can be transferred, e.g., to the \textit{musicology domain}, by adapting solutions to comply with the requirements of the existing models. \autoref{fig:method_transfer} outlines our methodology transfer model inspired by Simon et al.'s \textit{Liaison}~\cite{DBLP:conf/vissym/SimonMKS15}. Depending on how well the original solution aligned its problem with its design and on how similar the problems in both domains are, the design has to integrate more or fewer adaptations. The overall target of applying the MTM is to enlarge the solution space through available knowledge of every domain that is suitable for realizing methodology transfer.

\comment{
\subhead{Positioning}
\todo{Already covered at end of intro?}
We address the still existing gap of potential collaboration between musicology and visualization researchers to nourish future synergies. Both research domains can benefit from each other: Visualization researchers can create new methods by finding solutions for current unsolved tasks of musicologists. On the other side, musicology problems can be addressed by tailoring previously proposed visualization design models to answer musicology-related questions. Through this position paper, we want to motivate both sides to start cooperation. 
To facilitate such collaborations, we provide a description of the \textit{problem} and \textit{design space} of musicology and present a few uses cases to show gaps for future research opportunities. In Section~\ref{usecases}, we show exemplarily the success of already conducted methodology transfer from other domains to musicology.
\begin{itemize}
    \item MusicVis
    Typical Features: . Meta information: Composers and their works and lives. 
    \item SoS: TextVis \cite{DBLP:journals/computers/AlharbiL19}
    \item TimeViz Browser~\cite{timevizbrowseraigner2015}
    Examples of time visualization: Finance Data (Stock market). Temporal Visualization. 
    \begin{itemize}
        \item Data Features -- Frame of Reference $\rightarrow$ Abstract, Spatial  
        \item Data Features -- Number of Variables $\rightarrow$ Univariate, Multivariate  
        \item Time -- Arrangement $\rightarrow$ linear, cyclic
        \item Time -- Time Primitives $\rightarrow$ Instant, Interval
        \item Visualization -- Mapping $\rightarrow$ Static / Dynamic
        \item Visualization -- Dimensionality $\rightarrow$ 2D, 3D
    \end{itemize}
    \item High-Dimensional (BioVis~\cite{kerren_biovis_2017})
    \item Methodology Transfer based on Liaison~\cite{DBLP:conf/vissym/SimonMKS15}
\end{itemize}
}
\section{Visual Musicology}
We define the term \emph{Visual Musicology} to cover all characteristics of the musicology and visualization domains that are relevant to building \textit{Visual Musicology} systems. To be able to conduct a methodology transfer as described in \autoref{sec:methodology}, we first need to span the problem and design spaces of \textit{Visual Musicology}. 

In general, any visualization problem depends on the combination of domain users, data types, and tasks that users want to accomplish~\cite{DBLP:journals/cg/MikschA14}. \autoref{fig:teaser} displays the core components of the \textit{Visual Musicology} problem and design spaces, giving examples for each branch. These branches provide an in-depth description of the musicological interface for transferring to and from other domains. 
The problem space of \textit{Visual Musicology} is described by the different application \domain{Domains}, and their corresponding \data{Data}. For the domain branch, we consulted the research tracks suggested by the Journal of Interdisciplinary Music Studies (JIMS)~\cite{jims2019}. For the data branch, based on initial feedback from a musicology expert, we extracted exemplary types of data that are to be analyzed in each of the different application domains of the graph.
The design space of \textit{Visual Musicology} is split into the design for \analytic{Analytic Tasks} and for \visual{Visualization Tasks}. For the visualization branch, we adopted the visualization tasks from Kucher and Kerren~\cite{DBLP:conf/apvis/KucherK15}, which are partly based on  Shneiderman's information visualization mantra~\cite{shneiderman_eyes_1996}. For the analytic branch comprised the topics covered by Weihs book on music data analysis~\cite{weihs_music_2017}.
The solution space will be explored according to the methodology transfer (\autoref{sec:methodology}) from previously established domain definitions such as time-oriented visualizations by Aigner et al.~\cite{DBLP:series/hci/AignerMST11} as presented in the ``TimeViz Browser''~\cite{timevizbrowseraigner2015} and text visualization by Alharbi and Laramee~\cite{DBLP:journals/computers/AlharbiL19} or by Kucher and Kerren's~\cite{DBLP:conf/apvis/KucherK15}.

In the following, we give more detailed descriptions of the problem and design space of \textit{Visual Musicology} and conduct the methodology transfer (\autoref{sec:methodology}) for the Text, Geo, Time and HD domains. 

\begin{figure*}[t]
    \centering
    
\begin{annotatedFigure}
    {\includegraphics[width=\linewidth]{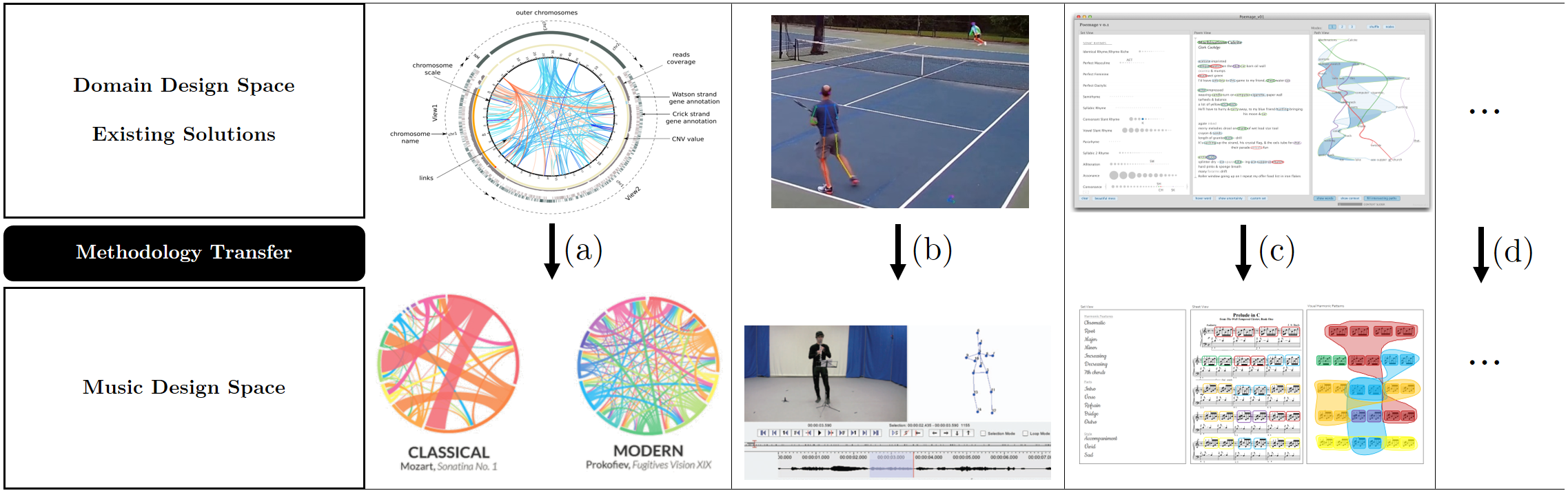}}
	\annotatedFigureText{0.23,0.92}{black}{0.31}{\cite{DBLP:journals/bmcbi/NaquindTS14}}
	\annotatedFigureText{0.23,0.35}{black}{0.31}{\cite{kellyschroerharmony2019}}
	\annotatedFigureText{0.465,0.92}{black}{0.31}{}
	\annotatedFigureText{0.465,0.35}{black}{0.31}{\cite{desmet_assessing_2012}}
	\annotatedFigureText{0.675,0.92}{black}{0.31}{\cite{DBLP:journals/tvcg/McCurdyLCM16}}
\end{annotatedFigure}
    \caption{
    Exemplary use cases for Methodology Transfer from different domains into Visual Musicology: (a) Exploiting the Circos visualization to identify harmonic relationships. (b) Analysis of embodied music interaction: Transferring computational methods of movement in tennis to elicit new information from musicians during performance. (c) Mapping the idea of analyzing the typology of poems from Poemage~\cite{DBLP:journals/tvcg/McCurdyLCM16} to analyzing the structure of musical features in music pieces.
    }
    \label{fig:methodusecases}
\end{figure*}

\subsection{Problem Space}
The \textit{problem space} comprises many different musicology challenges such as getting insight into complex harmonic relationships or understanding the relationship between body movements and music also called ``Embodied Music Cognition''~\cite{leman2008embodied}. There are manifold subdomains of musicology listed in the Journal of Interdisciplinary Music Studies~\cite{jims2019}. Their data ranges from structured information, such as musical features represented in a music sheet (e.g., \data{Pitch}), which could be analyzed, e.g., in the subdomain \domain{Theory \& Analysis}, to complex collections of multiple different signals (e.g. \data{Audio Recording}), which could be analyzed, e.g., in the subdomain \domain{Performance}. Some subdomains, such as \domain{Psychology} or \domain{Physiology, Medicine \& Therapy}, focus on the immediate effect of music on an individual or an \data{Audience}, while other subdomains, such as \domain{History, Cultural Studies \& Ethonomusicology} or \domain{Popular Culture \& Sociology}, focus on the long-term effect of music on culture and society at certain \data{Geographic} locations or within certain \data{Epochs}. Finally, some domains, such as \domain{Education} and \domain{Composition}, need to draw on a combination of data sources to generate carefully balanced solutions.

\subsection{Design Space}
Every domain has its own design space and may overlap with those of other domains. Using an abstract representation with \visual{Visualization Tasks} and \analytic{Analytic Tasks} helps to identify overlaps or commonalities between different domains and how to map two design spaces from different domains. In \autoref{fig:method_transfer}, the green circle-star connections indicate established problem solution mappings, while the stars in blue represent visualization techniques or models that have not been identified in a respective domain. To establish an idea of the visual musicology design space, we want to discuss some existing solutions. For instance, Sapp's keyscape visualization~\cite{sapp_visual_2005} is a design that enables the analysis of \analytic{hierarchical harmonic patterns in music pieces}. The visual representation is targeting both a quick \visual{Overview} and an efficient \visual{Navigation} to specific \visual{Regions of Interest}. Malandrino et al.~\cite{DBLP:conf/iv/MalandrinoPZZ15} proposed an application to support \analytic{structural analysis} of harmony in musical compositions on a more detailed level. Their visualization targets the \visual{Comparison} and \visual{Classification} of \visual{Regions of Interest} based on their harmonic structure. Many applied techniques are general and have been used in other domains, such as Wattenberg's arc diagram for text. Others are tailored to domain-dependent constraints, such as Malandrino's harmonic structures~\cite{DBLP:conf/iv/MalandrinoPZZ15}. The visual and analytic mapping in the design space underlies many that need to be considered to generate effective solutions. 

\subsection{Solution Space}
The solution space encloses mappings between the problem and design space. We will provide several examples that describe how existing solutions of different fields (e.g., text visualization)  can be transferred to solve musicological problems. To achieve this \textbf{methodology transfer}, we start by choosing a musicological application domain such as \domain{Theory \& Analysis} (see \autoref{fig:teaser}). Then we analyze the available musicological input data to detect a suitable visualization. During this process, we must take domain-specific constraints into account that might differ from other domains. The result is the basis for providing answers to the questions of the selected problem. This process should conclude with an evaluation of how well the solution deduced from the methodology transfer fits the initial problem. 

\subhead{Text Visualization --}
How can we identify text domain solutions that are applicable to music? For example, both text and music data have many structural features. If we take a look to the branch \domain{Domain} in \autoref{fig:teaser}, we can select  \domain{Theory \& Analysis} since it primarily deals with the structural features of data. 
Text and literature analysis are closely related to music theory and analysis. Based on this analogy, we start to find appropriate mappings of the input data between the domains. The musical property \data{Rhythm} comes close to accentuation or reading flow of text. We map {notes} with a certain \data{Pitch} to single letters, while chords represent whole words.~\data{Timbre} describes the sound or ``color'' of an instrument the influences the conceived sound. The corresponding aspect for text is an author's or speaker's characteristic writing style. We consider \data{Dynamics} as the emphasis of single words, accentuation, sentence types, or expressions. If required, we can further extend these mappings. After choosing a selection, we can apply, for instance, a \analytic{Structural Analysis} that extracts typical patterns that reflect the writing style of an author. Similarly, the same method provides typical patterns of composers showing their individual style. \mbox{El-Assady} et al.~\cite{DBLP:conf/acl/El-AssadyHGBHK17} implemented a visual text analytics framework to support deliberation analysis and the identification of patterns and speakers in debates. In this example, the visualization tasks \visual{Overview} and \visual{Exploration} are executed.

\subhead{Geo-Spatial Visualization --} 
A major goal of geo-visualizations is to understand geo-spatial relationships in data. \autoref{fig:teaser} lists \data{Geographic} as a data attribute in the sub-branch \textit{Meta-Information} of data. Also, \data{Epoch} and \data{Genre} are musicological features with spatial features.
To apply a transfer for these data types, we select the musicological field of \domain{History, Cultural Studies \& Ethnomusicology} of the branch \domain{Domain} in \autoref{fig:teaser}. The change of genres depends on the epoch and the geographic position of composers. For instance, different composers of various epochs followed different composition styles. These genres changed over time and were influenced by cultural circumstances and the invention of new instruments. Based on meta-information, musicologists can benefit from visualization tasks such as \visual{Exploration}, \visual{Region of Interest}, and \visual{Uncertainty Handling} to retrieve an overview of where and in which epoch composers lived an worked. Suitable visualizations with search parameters would enable musicologists to understand the relationships between composers as well as who might have had an influence on the composition style of another composer. Jänicke et al.~\cite{DBLP:journals/tvcg/JanickeFS16} proposed a VA system to enable  a \analytic{contextual analysis} (confession, occupation, location) of musicians from different epochs.

\subhead{Time Visualization --}
Time visualization covers the analysis of time-dependent data. In this exemplary transfer, we want to focus on the musicology application domain of \domain{Performance} from the branch \domain{Domain} in \autoref{fig:teaser}. In the analysis of performance, the time passing is a central variable. Two important aspects targeted by time-oriented visualization are the detection of recurring patterns and the comparison of patterns and their interdependence over time. In the case of performance analysis, this might be the recurrence of \data{gestures} the performing artist makes during one concert. The comparison of different timelines could be useful for the analysis of group performances with different performers or the analysis of the dependencies between changes in the \data{audio recording} and changes in the body movement of the performer. Once the performance data is visualized, the musicologist can use the visualization tasks \visual{Comparison} and \visual{Regions of Interest} to \analytic{analyze the characteristics of one performance} or multiple performances of one artist. One example of an existing tool that could be used to analyze performance data is Chronolenses~\cite{zhao2011exploratory}.

\subhead{High-Dimensional Visualization --}
The community of high-dimensional visualizations has a wide area of application. In this exemplary transfer, we want to focus on the musicology application domain of \domain{Physiology, Medicine \& Therapy} from the branch \domain{Domain} in \autoref{fig:teaser}. One area of HD visualization with a close correlation to music therapy is the analysis in psychological studies, e.g. ecological momentary assessment of behavior~\cite{shiffman2008ecological}. Regarding psychology, the momentary reaction would be specific behaviour as a result of external intervention. In the case of music therapy, the analogous assessment would be the momentary audience's \data{Perception} of and reaction to the music-based intervention. Both cases require a visual \visual{Exploration} of the measured data, including the assessment of multiple interdependent measurement sequences, the context, and different participant attributes. The participant meta-data, such as \data{Preference} for music and personality type for psychology is required as an aggregator for the analysis of patterns from multiple participants. Concerning the goals of the visual analysis, both cases want to find patterns in the effect of their intervention/music (\analytic{Perception Analysis}) on the participant and to relate them to personal/contextual factors to create reproducible behavioural theories/therapy practices. One example of a high dimensional visualization for behaviour that could be transferred to data retrieved from music therapy is Chronodes~\cite{DBLP:journals/tiis/PolackCKBBSC18}.
\section{Use Cases}
\label{usecases}
We provide three different use cases of applying methodology transfer to derive a visual musicology solution. 

\subhead{(a) Visual Analysis of Harmony  -- } 
Schroer~\cite{kellyschroerharmony2019} exploited the radial layout of the circos graph~\cite{DBLP:journals/bmcbi/NaquindTS14} to visualize the chord harmony in their respective tonal system of Western Art Music. Originally, Naquin et al. proposed the circos graph to enable visual analysis of structural genome variations. Therefore, Schroer got inspired by a visualization for biological data to display dominant chord relationship by mapping tonal degrees to color. Figure~\ref{fig:methodusecases}(a) shows that multiple adaptations were applied to the original model, but the general idea is preserved. The results for single pieces of different epochs show that there is a dominant chord in classical music (e.g., Mozart), which is highlighted in red. In contrast, the exemplary modern music piece by Prokofiev contains an almost equal distribution of every degree. Consequently, this use case is an example of a successful mapping of an available problem - design space mapping.
    
\subhead{(b) Embodied Music Interaction  -- } 
Current research on \textit{Embodied Music Interaction}~\cite{leman_expressive_2016,lesaffre_routledge_2017} faces the challenge to understand the complex relationship between music, performers, audience, and contexts. These aspects have the characteristics of a complex dynamic system having both stable and in-between interaction states that change over time. \autoref{fig:methodusecases}(b) shows how automatic analysis of video footage allows the recognition of the human body to analyze the performance in tennis. Applying such methods to the body movement of a clarinet player~\cite{desmet_assessing_2012} could enable the identification of recurring patterns of musicians' gestures. Music scores can be conceived as instruction sets for human actors whose gesturing underlies music playing. Of course, this use case requires looking at audio data which might not be relevant in tennis. Therefore, an additional component that supports such analysis tasks is necessary to relate body movement data with performance recordings.
    
\subhead{(c) Investigating Musical Features -- } 
\textit{Poemage} is a user interface to visualize the sonic topology of poems~\cite{DBLP:journals/tvcg/McCurdyLCM16}. Linguists and other digital humanities researchers who analyze textual data can use Poemage for \textit{close reading} to retrieve in-depth information about poems. With this application, text analysts can execute an exhaustive text investigation with the support of automatic highlighting and semantic relationships, such as rhymes by the use of sound in poetry. Visualizations help reveal hidden relationships through the employment of well-polished designs to cope with the challenges of textual close reading. We argue that Poemage is a suitable source of inspiration to apply similar or equal approaches on single music documents to enable \textit{close reading}~\cite{janicke_close_2015} based on different musical features such as the role of certain chords in their context or modulation. \autoref{fig:methodusecases} (c) indicates how the visual interface of Poemage can be adjusted to visualize the \textit{harmonic topology of a piano composition}, for instance. Detail views and colored highlighting can be employed to emphasize interesting harmonic progressions, and a feature panel on the left could provide analysts with powerful tools to support in-depth analysis of musical pieces. Of course, the application must be adapted for musical data, and different methods to extract relevant information from the piece have to be integrated. Nevertheless, this application transfer could motivate music researchers to use such an application, since a similar approach has been successfully introduced to another digital humanities community. 
    
We emphasize that there is still much potential for numerous new methodology transfer applications (see Figure~\ref{fig:methodusecases}(d)). To identify such research opportunities, close collaboration with domain experts and musicologists is required~\cite{DBLP:conf/vissym/SimonMKS15}. For instance, use case (b) describes a scenario that has not been applied yet and may serve as a starting point to apply body movement based on skeleton analysis models to the movement and gestures of musicians during performances. Of course, factors like cultural context and biological aspects are relevant for understanding the dynamics of music interaction such as in higher-level music training ~\cite{caruso_gestures_2016,antoniadis_processing_2016}, interactive multimedia art~\cite{vets_gamified_2017,maes_embodied_2018}, sports~\cite{lorenzoni_design_2018}, and medical rehabilitation\cite{moumdjian_entrainment_2018}. 



\section{Research Implications and Opportunities}
Our research introduces how musicologists and visualization researchers can initiate collaboration by applying \textit{methodology transfer} for \textit{Visual Musicology}. Furthermore, \textit{Visual Musicology} provides countless opportunities for original research. In summary, we make seven claims designed to motivate musicologists and visualization researchers alike to increase mutual collaboration.

\textbf{(1)~Visual Musicology is more than visualizing music data -- }
There is more to Visual Musicology than merely visualizing music. \autoref{fig:teaser} lists multiple musicology domains covering a diverse range of data characteristics and analytic tasks. For instance, psychological, physiological, and philosophical considerations are required to address relevant musicology topics and to understand the complexity of research questions in this domain. Often, data analysts only apply structural analysis on music data which we consider to be only a subfield of visual musicology reflected by \domain{Theory \& Analysis} and \domain{Computing} as part of the domain branch in \autoref{fig:teaser}.

\textbf{(2)~Visual Musicology is an under-researched field --}
Multiple challenges and problems that musicologists are facing still require special attention to find suitable solutions~\cite{muller_computational_2016}. But even for existing solutions, information visualization research could lead to more effective and efficient results. For example, during Dagstuhl Seminar 16092, numerous possible research ideas regarding \textit{Computational Music Structure Analysis} were discussed~\cite{muller_computational_2016}. Musicology provides an extensive collection of diverse data sources. Different data analysis techniques such as machine learning, e.g., with probabilistic models, and human-based analysis methods, e.g., with visual analytics, can be applied and evaluated using these data sources.

\textbf{(3)~Musicology is expanding through technological progress --} 
Technological advances, such as the development of virtual reality environments, the creation of new instruments, and the influence of music from different cultures due to the access of the internet, lead to new considerations of musicological questions. The possibilities of gathering data from brain activity in neuroscience enable research in different domains, which were not possible decades ago. Psychological, physical, and physiological factors are interesting aspects of musicologists. Automatic generation of music with artificial intelligence is on the rise and moving into focus within musicology. Employing visualization to facilitate understanding of such complex relationships can lead to a better comprehension of musicology topics. The technological progress expands musicology, raises new research questions, opens new data sources, and requires new solutions.

\textbf{(4)~Musicology offers rich collections of data -- }
The emergence of improved digital format standards such as MusicXML~\cite{good_musicxml_2001}, the invention of new instruments, and the increased usage of streaming music services such as Spotify or YouTube provide large data sources. Also, big data collections like IMSLP~\footnote{https://imslp.org/}, ``The Million Song Dataset''~\cite{DBLP:conf/ismir/Bertin-MahieuxEWL11}, BMLO~\footnote{http://www.bmlo.lmu.de/}, or MUSICI and MusMig~\cite{over_musici_2016} are available today. Such data sources not only provide music data, but often additional information about the life and work of (early) composers, their relationships, or genres. The new amount and diversity of available data poses challenges to musicology but also opens up chances for interdisciplinary solutions such as the application of visual analytics.


\textbf{(5)~Complex musicological data requires analytic and visual support  -- }
Musicologists could benefit from assistance through visual and analytic solutions to cope with domain challenges. Conducting manual analysis or studies on such extensive datasets is often not feasible. Thus, musicologists are not able to realize the full potential of their datasets without corresponding computational approaches and strong interdisciplinary collaboration. For instance, there are experimental musicological scenarios such as embodied music cognition~\cite{leman2008embodied} or embodied music interaction~\cite{maes_embodied_2018} in which domain experts need to deal with complex body movement data. Adapting available visualization techniques could reveal previously unknown patterns and enable the knowledge generation for the musicology domain.

\textbf{(6)~Applying methodology transfer can open up new fields of interdisciplinary research -- }
We propose a methodology transfer model (MTM) (see \autoref{sec:methodology}) to present open challenges and motivate interdisciplinary research. We postulate that visualization research is a valuable source to support digital humanities researchers. Similarly, musicologists have the required domain knowledge to guide problem-oriented collaboration by providing data and real-world applications. Implementing the MTM instead of creating solutions from scratch saves precious research work. The MTM also allows focusing on issues that require new solutions due to the identification of research gaps during the methodology transfer process.

\textbf{(7)~The digital humanities (DH) offer valuable insights for information visualization research -- } 
Visualization researchers typically work on methodologies to be applied to the problems of other domains. 
However, due to the experience and long history of humanities, there is much value in applying methodologies from  DH research  for information visualization~\cite{wrisley_visualization_2018}. While other domains benefit from close cooperation with visualization experts, they should not hesitate to learn from the humanities' research experience of many centuries. Instead of considering the DH only as data providers, we should start considering them as equal partners with different views on the world that we should learn from~\cite{BEC+18}.

\comment{
\begin{itemize}
    \item We have discussed different opportunities of future work.
    \item Existing research gap indicate potential interdisciplinary research work and both parties can benefit from collaboration
    \item Solution Space Continuum: Different constraints require more or less work to tailor existing visualization design models to serve as a solution for a musicological issue
    \item Different data characteristics indicate where to search for existing methods to apply on musicological problems: Geo, Temporal, Structural, Pattern Mining, Machine Learning, Artificial Intelligence
    \item Learn from existing domains as text where much more research work have been done already
    \item In the domain of visual musicology there are many issues that do not have a solution and require individual attention. Developing new methods could yield new visualization techniques and are anchored to be a solution for musicologists.
    \item \textcolor{red}{\textbf{@Marc:} High demand in centers like Marc's to visualize data collected during performance + Example and concrete references to related projects, e.g., project in Barcelona.\\ Visualization of multiple sources of data, challenges on how to represent such a multi-modal resource.}
\end{itemize}
}






\section{Conclusion}
We propose how to utilize available visualization techniques and applications with musicology. Thus, we frame \textit{visual musicology} by providing a diligent overview of substantial musicological domains such as ethnomusicology or education. We describe typical computational domain tasks and data characteristics to define the design and problem space of visual musicology. We emphasize that many musicology problems can be addressed by tailoring existing visualization techniques to support musicologists in fulfilling their tasks. We see high potential for mutual benefits of such a cooperation since musicology provides many complex issues that require innovative solutions. We explain how \textit{methodology transfer} can be strategically executed to profit from state-of-the-art solutions. Eventually, motivated visualization researchers and musicologists alike are needed to develop and generate new approaches and knowledge through close interdisciplinary collaboration.

\comment{
\begin{itemize}
    \item methodology transfer
    \item Visual Musicology
    \item  emphasize the potential for fruitful collaborations between music and visualization researchers
    \item presented how existing visualization techniques can be applied
\end{itemize}
}

\newpage

\bibliographystyle{abbrv-doi-hyperref}
\bibliography{_references}
\end{document}